\documentclass[11pt]{article}
\usepackage[utf8]{inputenc}
\usepackage[colorlinks=true,citecolor=blue,linkcolor=blue]{hyperref}
\usepackage{multirow}
\usepackage[left=2cm,right=2cm,top=2cm,bottom=2cm]{geometry}
\usepackage{cite}
\usepackage{psfrag}
\usepackage{graphicx}
\usepackage{epsfig}
\usepackage[english]{babel}
\usepackage{booktabs}
\usepackage{amsmath,amsfonts,amssymb}
\usepackage{pstcol,pst-fill,pst-grad}
\usepackage{pstricks,pst-fill,pst-grad}
\usepackage{euscript}
\usepackage{pstricks}
\usepackage{wrapfig}
\usepackage{slashed}
\usepackage[toc,page]{appendix}
\usepackage{here}

\date{}
\begin{document}
	\title{\vspace{-3cm}
		\hfill\parbox{4cm}{\normalsize \textit{}}\\
		\vspace{1cm}
		{Laser-assisted charged Higgs boson decay in Two Higgs Doublet Model - type II}}
	\vspace{2cm}
	\author{S. Mouslih$^{1,2}$, M. Jakha$^1$, S. El Asri$^1$, S. Taj$^1$, B. Manaut$^{1,}$\thanks{Corresponding author, E-mail: b.manaut@usms.ma}, R. Benbrik$^3$ and E Siher$^2$\\
		{\it {\small$^1$ Polydisciplinary Faculty, Laboratory of Research in Physics and Engineering Sciences, }}\\
		{\it {\small  Team of Modern and Applied Physics, Sultan Moulay Slimane University, Beni Mellal, 23000, Morocco.}}
		\\		
	{\it {\small$^2$Faculty of Sciences and Techniques,
		Laboratory of Materials Physics (LMP),
		Beni Mellal, 23000, Morocco.}}		
		\\			
	{\it {\small$^3$ Polydisciplinary Faculty, Laboratoire de Physique Fondamentale et Appliquée, Sidi Bouzid, B.P. 4162 Safi, Morocco.}}		
	}
\maketitle \setcounter{page}{1}
\date{\today}

\begin{abstract}
In this paper, we investigate the charged Higgs boson decay in the context of the type-II two-Higgs-doublet model in the presence of a circularly polarized electromagnetic field of laser radiation. The calculations are performed by adopting the Furry picture approach of non-perturbative interactions with the external electromagnetic field. Using the method of exact solutions for charged particles states in the presence of a circularly polarized electromagnetic wave field and evaluating the $S$-matrix elements, an exact analytic expression is derived for the decay width of leptonic, hadronic and bosonic decay modes. The branching ratios of different decay modes with multiple photon emission and absorption from the laser beam are analyzed and found to be dramatically modified in the region of superstrong fields. The dependencies of the decay width on the laser field strength and frequency are also examined. The results obtained may be interesting for future experimental and theoretical investigations.
\end{abstract}

\section{Introduction}
The discovery of a Higgs boson in the mass region of about $125$ GeV, announced by ATLAS \cite{atlas} and CMS \cite{cms} at the Large Hadron Collider (LHC) in the summer of $2012$, filled a gap in the Standard Model (SM), the theory that describes all the particles and interactions that make up the universe. However, despite its good agreement with the experimental data available today, it is still too early to consider the SM as the ultimate theory of particle interactions. Therefore, the need to explore extensions beyond the SM is obvious. The Two-Higgs-Doublet Model (2HDM), where just a second scalar Higgs doublet is added to the already existing one in the SM, is one of the simplest and minimal extensions beyond SM \cite{2hdm}. Its scalar sector contains five physical states: a light CP even neutral Higgs $h$, a heavy CP even neutral Higgs $H$, a CP odd (pseudoscaler) neutral Higgs $A$ and a pair of charged Higgs $H^\pm$, with $h$ being the SM-like Higgs boson observed at LHC. Depending on which type of fermions couples to which doublet, 2HDMs can be categorized into different types. Currently, the main focus is on the so-called type-II 2HDM \cite{thesis}, because it is an essential feature of the Minimal Supersymmetric Standard Model (MSSM). One of the most striking signs of physics beyond SM would be the appearance of a charged Higgs boson. However, the prospects for hunting the charged Higgs are quite difficult \cite{prospect1,prospect2}. Recent searches for $H^+$ at the LHC focus on production and decay via interactions with SM fermions. Such high-energy physics experiments require increasingly powerful and high-energy particle collisions. In view of the rapid development of modern laser devices \cite{laser1,laser2,laser3}, laser acceleration becomes more and more interesting and can be a promising solution to increase the required collision energy \cite{donald1999}. In this context, it is very motivating to study the processes of Higgs production and decay in the presence of strong laser fields. In 2014, Sarah Muller et al. \cite{muller1,muller2} conducted pioneering research on Higgs boson production and various particle physics processes in lepton collisions boosted by a laser field. Recently, another group of authors has also studied different physical processes in the SM and beyond \cite{bsm1,bsm2,bsm3,bsm4,bsm5}, including charged Higgs boson production \cite{hplus1,hplus2,hplus3}. A review of such contributions in strong field quantum electrodynamics (QED) can be found in \cite{qed1,qed2}. Moreover, laser-assisted electroweak decay processes have also recently attracted the attention of many researchers and have been the subject of numerous papers, among which we refer to \cite{decay1,wdecay,zdecay,decay3,decay4,decay5}. The purpose of the present study is to consider the laser-assisted charged Higgs decay within the framework of 2HDM type-II.  This is done by addressing the three decay modes; namely, leptonic, hadronic and bosonic. Our theoretical calculation is based on the Furry picture approach \cite{furry}, where we evaluate the lowest-order $S$-matrix elements by using Volkov solutions \cite{volkov} to describe charged particles inside the laser field. The rest of the paper is organized as follows. First, we establish the expressions of decay width for different decay modes in Sect.~\ref{sec:theory}. In Sect.~\ref{sec:res}, we discuss the numerical results obtained, and we finally conclude in Sect.~\ref{sec:conclusion}. Note that natural units $c=\hbar=1$ are used throughout.
\section{Outline of the theory}\label{sec:theory}
We consider the decay of the charged Higgs boson $H^+$ in the presence of a laser field. Since we have three decay modes, this section is divided into three subsections. Each one is dedicated to the calculation of the decay width for each decay mode. The calculation is detailed for the first decay mode, and elsewhere we limit ourselves to the necessary and give the final expressions. We will start with the leptonic decay, then the hadronic and finally the bosonic decay mode.
\subsection{Leptonic decay width of the charged Higgs in the presence of a laser field}
\begin{figure}[hbtp]
\centering
\includegraphics[scale=0.6]{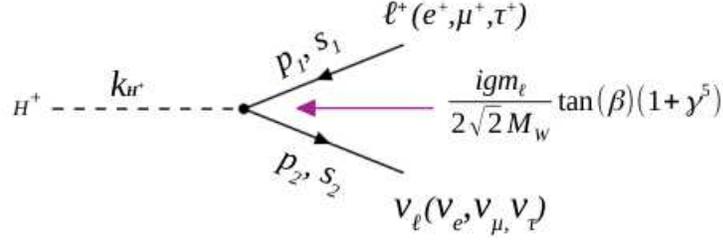}
\caption{Feynman diagram for the leptonic decay of the charged Higgs boson ($H^{+}\rightarrow \ell^{+}\nu_{\ell}$).}\label{diagram1}
\end{figure}

We begin by considering the leptonic decay process of charged Higgs boson $H^{+}$, with 4-momentum  $k_{H^+}$ and mass $M_{H^+}$, into an antilepton ($\ell^+$) and a corresponding neutrino ($\nu_{\ell}$) in a circularly polarized electromagnetic (EM) field. It can be schematized as follows:
\begin{equation}\label{processus}
H^{+}(k_{H^{+}})\longrightarrow \ell^{+} (p_{1})+\nu_{\ell}(p_{2}),~~~~\ell\equiv (e,\mu,\tau),
\end{equation}
where $p_{1}$ and $p_{2}$ are the free 4-momenta of antilepton and neutrino. The corresponding Feynman diagram is presented in Fig.~\ref{diagram1}.
We take the laser field as a circularly polarized monochromatic EM field, which is classically described by the following 4-potential
\begin{align}\label{potential}
A^{\mu}(\phi)=a^{\mu}_{1}\cos(\phi)+a^{\mu}_{2}\sin(\phi),
\end{align}
where $\phi=(k.x)$ is the phase of the laser field. $k=(\omega,\textbf{k})$ is the wave 4-vector and $\omega$ is the laser frequency. The  4-amplitudes $a^{\mu}_{1}=|\textbf{a}|(0,1,0,0)$ and $a^{\mu}_{2}=|\textbf{a}|(0,0,1,0)$ are orthogonal and equal in magnitude, which implies $(a_{1}.a_{2})=0$ and $a_{1}^{2}=a_{2}^{2}=a^{2}=-|\textbf{a}|^{2}=-(\xi_{0}/\omega)^{2}$ where $\xi_{0}$ is the electric field strength. We assume that this 4-potential satisfies the Lorentz gauge condition, $k_{\mu}A^{\mu}=0$, which means $(k.a_{1})=(k.a_{2})=0$, indicating that the wave vector $\textbf{k}$ is chosen to be along the $z$-axis. \\
The equivalent low-order $S$-matrix element for the process (\ref{processus}) in 2HDM type II can be written as follows \cite{greiner}:
\begin{equation}
S_{fi}=\frac{ig m_{\ell}}{2\sqrt{2}M_{W}}\tan(\beta)\int d^{4}x\overline{\psi}_{\nu_{\ell}}(x,p_{2})(1+\gamma^{5})\psi_{\ell^+}(x,p_{1})\overline{\phi}_{H}(x,k_{H^{+}}),
\end{equation}
where $g$ is the electroweak coupling constant. $m_{\ell}$ and $M_{W}$ are the mass of the outgoing lepton and $W$-boson. In order to take into account the interaction of the charged Higgs (0-spin particle) with the laser field, its wave function obeys the Klein-Gordon equation for bosons of spin zero. Hence, the corresponding Volkov solutions are given, normalized to the volume $V$, as follows \cite{Szymanowski}:
\begin{equation}\label{Higgswave function}
\phi_{H}(x,k_{H^{+}})=\frac{1}{\sqrt{2QV}}\times e^{iS(q,x)},
\end{equation}
where $Q$ is the total energy of $H^+$ in the presence of the laser field, and
\begin{equation}
S(q,x)=q.x-\frac{ e(a_{1}.k_{H^{+}})}{(k.k_{H^{+}})}\sin(\phi)+\frac{ e(a_{2}.k_{H^{+}})}{(k.k_{H^{+}})}\cos(\phi),
\end{equation}
where $e=-|e|<0$ is the charge of electron, and the 4-momentum $q$ of the dressed $H^+$ is given by
\begin{equation}
q_=k_{H^{+}}-\frac{e^{2}a^{2}}{2(k.k_{H^{+}})}k.
\end{equation}
The square of $q$ reads
\begin{equation}\label{meffH}
q^{2}=M^{*2}_{H^{+}}=M_{H^{+}}^{2}-e^{2}a^{2},
\end{equation}
where $M^{*}_{H^+}$ is the effective mass of $H^+$ acquired inside the laser field.\\
The outgoing antilepton is described by the relativistic Dirac-Volkov functions \cite{volkov} normalized to the volume $V$ as follows:
\begin{equation}\label{hwave function}
\psi_{\ell^+}(x,p_{1})=\bigg[1-\frac{ e\slashed{k}\slashed{A}}{2(k.p_{1})}\bigg]\frac{v(p_{1},s_{1})}{\sqrt{2Q_{1}V}}\times e^{iS(q_{1},x)},
\end{equation}
where
\begin{equation}
S(q_{1},x)=q_{1}.x-\frac{ e(a_{1}.p_{1})}{(k.p_{1})}\sin(\phi)+\frac{ e(a_{2}.p_{1})}{(k.p_{1})}\cos(\phi).
\end{equation}
$v(p_{1},s_{1})$ is the free Dirac spinor for the antilepton with 4-momentum $p_{1}$ and spin $s_{1}$ satisfying the following relation: $\sum_{s_{1}}v(p_{1},s_{1})\overline{v}(p_{1},s_{1})=\slashed{p}_{1}+m_{\ell}$. The 4-vector $q_{1}=(Q_{1},\textbf{q}_{1})$ is the quasi-momentum, which is related to $p_1$ by the following relation:
\begin{equation}
q_{1}=p_{1}-\frac{e^{2}a^{2}}{2(k.p_{1})}k,~~~q_{1}^{2}=m^{*2}_{\ell}=m_{\ell}^{2}-e^{2}a^{2},
\end{equation}
where $m^{*}_{\ell}$ is the effective mass of the anti-leptons inside the laser field.\\
The outgoing neutrino, which is electrically neutral, does not interact with the EM field. Its wave function is given by \cite{greiner}:
\begin{align}\label{neutrinowave}
\psi_{\nu_{\ell}}(x,p_2)=\dfrac{u(p_{2},s_{2})}{\sqrt{2E_{2}V}}\times e^{-ip_{2}.x},
\end{align}
where $E_{2}=p_{2}^{0}=|\textbf{p}_{2}|$, and $u(p_{2},s_{2})$ is the free Dirac spinor satisfying $\sum_{s_{2}}u(p_{2},s_{2})\overline{u}(p_{2},s_{2})=\slashed{p}_{2}$. \\
After some algebraic manipulations, we find that $S_{fi}$ becomes
\begin{equation}\label{Sfileptons}
\begin{split}
S_{fi}(H^{+}\rightarrow \ell^{+}\nu_{\ell})=&\frac{igm_{\ell}}{2\sqrt{2}M_{W}}\frac{\tan(\beta)}{\sqrt{8QQ_{1}E_{2}V^{3}}}\int d^{4}x \overline{u}(p_{2},s_{2})\Big[\Big(1+\gamma^{5}\Big)\Big(1-\frac{ e\slashed{k}\slashed{A}}{2(k.p_{1})}\Big)\Big]v(p_{1},s_{1}),\\
& \times e^{i(q_{1}+p_{2}-q).x}\times e^{-iz_{\ell}\sin(\phi-\phi_{0})},
\end{split}
\end{equation}
where we have used the following transformation:
\begin{equation}
e^{i(S(q_1,x)-S(q,x))}=e^{i(q_{1}-q).x} \times e^{-iz_{\ell}\sin(\phi-\phi_{0})},
\end{equation}
with
\begin{equation}\label{argumentlepton}
z_{\ell}=e \sqrt{\bigg(\frac{a_{1}.p_{1}}{k.p_{1}}-\frac{a_{1}.k_{H^{+}}}{k.k_{H^{+}}}\bigg)^{2}+\bigg(\frac{a_{2}.p_{1}}{k.p_{1}}-\frac{a_{2}.k_{H^{+}}}{k.k_{H^{+}}}\bigg)^{2}},
\end{equation}
and
\begin{equation}
\phi_{0}=\arctan\bigg[\frac{(a_{2}.p_{1})(k.k_{H^{+}})-(a_{2}.k_{H^{+}})(k.p_{1})}{(a_{1}.p_{1})(k.k_{H^{+}})-(a_{1}.k_{H^{+}})(k.p_{1})}\bigg].
\end{equation}
After integration over the space-time and using the transformation that introduces Bessel functions
\begin{equation}
e^{iz\sin(\phi)}=\sum_{n=-\infty}^{+\infty} J_{n}(z) e^{in \phi},
\end{equation}
the $S$-matrix element can be written as a sum of ordinary Bessel functions, such that
\begin{equation}
S_{fi}(H^{+}\rightarrow \ell^{+}\nu_{\ell})=\sum_{n=-\infty}^{+\infty} \frac{igm_{\ell}}{2\sqrt{2}M_{W}}\frac{\tan(\beta)}{\sqrt{8QQ_{1}E_{2}V^{3}}}(2\pi)^{4}\delta^{4}(q_{1}+p_{2}-q-nk)\mathcal{M}^{n}_{fi}.
\end{equation}
The quantity $\mathcal{M}^{n}_{fi}$ is defined by
\begin{equation}
\mathcal{M}^{n}_{fi}=\overline{u}(p_{2},s_{2})\Big[\Big(1+\gamma^{5}\Big)\Big(b_{n}(z_{\ell})-C(p_{1})\slashed{k}\slashed{a}_{1}b_{1n}(z_{\ell})-C(p_{1})\slashed{k}\slashed{a}_{2}b_{2n}(z_{\ell})\Big)\Big]v(p_{1},s_{1}),
\end{equation}
where $C(p_{1})=e/[2(k.p_{1})]$ and the three coefficients $b_{n}(z_{\ell})$, $b_{1n}(z_{\ell})$ and $b_{2n}(z_{\ell})$ are expressed as a function of ordinary Bessel functions by \cite{landeau}
\begin{equation}
\begin{split}
   & b_{n}(z_{\ell})=J_{n}(z_{\ell})e^{in\phi_{0}},\\
   &b_{1n}(z_{\ell})=\frac{1}{2}\big[ J_{n+1}(z_{\ell})e^{i(n+1)\phi_{0}}+J_{n-1}(z_{\ell})e^{i(n-1)\phi_{0}}\big],\\
 	&b_{2n}(z_{\ell})=\frac{1}{2i}\big[ J_{n+1}(z_{\ell})e^{i(n+1)\phi_{0}}-J_{n-1}(z_{\ell})e^{i(n-1)\phi_{0}}\big].\\
\end{split}
\end{equation}
$z_{\ell}$ ($\ell$ stands for leptonic) is the argument of Bessel functions already defined in Eq.~(\ref{argumentlepton}) and $n$, called the order, is interpreted as the number of photons exchanged between the laser field and the decay process.\\
To calculate the partial decay width, $\Gamma(H^{+}\rightarrow \ell^{+}\nu_{\ell})$, we multiply the square $S$-matrix element, $|S_{fi}|^{2}$, by the density of final states, and divide it by the time $T$ and finally, we have to make the average on the initial spins and the sum on the final ones. Thus, the laser-assisted leptonic decay width is expressed as follows:
\begin{align}
\Gamma(H^{+}\rightarrow \ell^{+}\nu_{\ell})=\sum_{n=-\infty}^{+\infty}\Gamma^{n}(H^{+}\rightarrow \ell^{+}\nu_{\ell}),
\end{align}
where the individual partial decay width for each $n$, $\Gamma^{n}(H^{+}\rightarrow \ell^{+}\nu_{\ell})$, is defined by
\begin{align}
\Gamma^{n}(H^{+}\rightarrow \ell^{+}\nu_{\ell})=\frac{g^{2}m^{2}_{\ell}\tan^{2}(\beta)}{64M^{2}_{W}Q}\int \dfrac{d^{3}q_{1}}{(2\pi)^{3}Q_{1}}\int \dfrac{d^{3}p_{2}}{(2\pi)^{3}E_{2}}(2\pi)^{4}\delta^{4}(q_{1}+p_{2}-q-nk)|\overline{\mathcal{M}^{n}_{fi}}|^{2},
\end{align}
where
\begin{align}\label{sums}
|\overline{\mathcal{M}^{n}_{fi}}|^{2}=\sum_{s_{1},s_{2}}\Big|\overline{u}(p_{2},s_{2})\Big[\Big(1+\gamma^{5}\Big)\Big(b_{n}(z_{\ell})-C(p_{1})\slashed{k}\slashed{a}_{1}b_{1n}(z_{\ell})-C(p_{1})\slashed{k}\slashed{a}_{2}b_{2n}(z_{\ell})\Big)\Big]v(p_{1},s_{1})\Big|^{2}.
\end{align}
Performing the integration over $d^{3}p_{2}$ and using $\delta^{4}(q_{1}+p_{2}-q-nk)=\delta^{0}(Q_{1}+E_{2}-Q-n\omega)\delta^{3}(\textbf{q}_{1}+\textbf{p}_{2}-\textbf{q}-n\textbf{k})$, we find
\begin{equation}
\Gamma^{n}(H^{+}\rightarrow \ell^{+}\nu_{\ell})=\frac{g^{2}m^{2}_{\ell}\tan^{2}(\beta)}{64(2\pi)^{2}M^{2}_{W}Q}\int \dfrac{d^{3}q_{1}}{Q_{1}E_{2}}\delta^{0}(Q_{1}+E_{2}-Q-n\omega)|\overline{\mathcal{M}^{n}_{fi}}|^{2},
\end{equation}
with $\textbf{q}_{1}+\textbf{p}_{2}-\textbf{q}-n\textbf{k}=0$. Choosing the rest frame of $H^{+}$ and using $ d^{3}q_{1}=|\textbf{q}_{1}|Q_{1}dQ_{1}d\Omega_{\ell}$, we get
\begin{equation}
\begin{split}
\Gamma^{n}(H^{+}\rightarrow \ell^{+}\nu_{\ell})=&\frac{g^{2}m^{2}_{\ell}\tan^{2}(\beta)}{64(2\pi)^{2}M^{2}_{W}Q}\int \frac{|\textbf{q}_{1}|}{E_{2}}|\overline{\mathcal{M}^{n}_{fi}}|^{2}dQ_{1}d\Omega_{\ell}\delta^{0}\bigg(Q_{1}-Q-n\omega\\
&+\sqrt{\Big(n-\frac{e^{2}a^{2}}{2(k.k_{H^{+}})}\Big)^{2}\omega^{2}+Q^{2}_{1}-m^{*2}_{\ell}+2\omega\sqrt{Q^{2}_{1}-m^{*2}_{\ell}}\Big(\frac{e^{2}a^{2}}{2(k.k_{H^{+}})}-n\Big)\cos(\theta)}\bigg),
\end{split}
\end{equation}
where we have replaced $E_{2}$ by its expression (the square root) to show the $Q_{1}$ dependence inside the delta function. The remaining integral over $dQ_{1}$ can be solved by using the familiar formula \cite{greiner}:
\begin{align}
\int dyf(y)\delta(g(y))=\dfrac{f(y)}{|g'(y)|}\bigg|_{g(y)=0}.
\end{align}
Thus, we get
\begin{equation}
\Gamma^{n}(H^{+}\rightarrow \ell^{+}\nu_{\ell})=\frac{g^{2}m^{2}_{\ell}\tan^{2}(\beta)}{64(2\pi)^{2}M^{2}_{W}Q}\int  \frac{|\textbf{q}_{1}||\overline{\mathcal{M}^{n}_{fi}}|^{2}d\Omega_{\ell}}{Q+n\omega+\frac{Q_{1}\omega\cos(\theta)}{\sqrt{Q_{1}^{2}-m^{*2}_{\ell}}}\Big(n-\frac{e^{2}a^{2}}{2(k.k_{H^{+}})}\Big)},
\end{equation}
where $g^{2}=8G_{F}M_{W}^{2}/\sqrt{2}$, with $G_{F}=(1.166~37\pm0.000~02)\times10^{-11}~\text{MeV}^{-2}$ is the Fermi coupling constant.\\
The term $|\overline{\mathcal{M}^{n}_{fi}}|^{2}$ in (\ref{sums}) can be calculated by converting the sums over the spins into traces as follows:
\begin{equation}\label{trace1}
|\overline{\mathcal{M}^{n}_{fi}}|^{2}=\text{Tr}\big[\slashed{p}_{2}\Lambda^{n}(\slashed{p}_{1}-m_{\ell})\overline{\Lambda}^{n}\big],
\end{equation}
where
\begin{equation}
\begin{split}
&\Lambda^{n}=\big(1+\gamma^{5}\big)\big(b_{n}(z_{\ell})-C(p_{1})\slashed{k}\slashed{a}_{1}b_{1n}(z_{\ell})-C(p_{1})\slashed{k}\slashed{a}_{2}b_{2n}(z_{\ell})\big),\\
&\overline{\Lambda}^{n}=\big(1-\gamma^{5}\big)\big(b^{*}_{n}(z_{\ell})-C(p_{1})\slashed{a}_{1}\slashed{k}b^{*}_{1n}(z_{\ell})-C(p_{1})\slashed{a}_{2}\slashed{k}b^{*}_{2n}(z_{\ell})\big).
\end{split}
\end{equation}
The trace calculation is performed with the help of FeynCalc \cite{feyncalc1}. Refer to appendix \ref{appendix} for the detailed and explicit expression of $|\overline{\mathcal{M}^{n}_{fi}}|^{2}$.
\subsection{Hadronic decay width of the charged Higgs in the presence of a laser field}
In the framework of beyond standard model, we consider the hadronic decay of $H^+$ into a pair of quarks
\begin{equation}\label{processus_hadr}
H^{+}(k_{H^{+}})\longrightarrow q (p_{1})+\overline{q}'(p_{2}),
\end{equation}
where $q\equiv (u,c,t)$ and $q'\equiv (d,s,b)$. The overall coupling between $H^+$ and the quarks is defined as follows \cite{marin2004}:
\begin{equation}
\text{vertex}-H^{+}\text{-}q\text{-}\overline{q}'=\frac{igV_{qq'}}{2\sqrt{2}M_{W}}(A+B\gamma^{5}),
\end{equation}
where $A=m_{q'}\tan(\beta)+m_{q}\cot(\beta)$ and $B=m_{q'}\tan(\beta)-m_{q}\cot(\beta)$. $V_{qq'}$ is the element of the CKM matrix. The lowest-order $S$-matrix element of the hadronic decay of $H^{+}$ is described as follows:
\begin{equation}\label{hadr Sfi}
S_{fi}(H^{+}\rightarrow q\overline{q}')=\frac{ig V_{qq'}}{2\sqrt{2}M_{W}}\int d^{4}x\overline{\psi}_{q}(x,p_{1})(A+B\gamma^{5})\psi_{\overline{q}'}(x,p_{2})\overline{\phi}_{H}(x,k_{H^{+}}),
\end{equation}
where $\psi_{q}$ and $\psi_{\overline{q}'}$ are, respectively, the relativistic Dirac-Volkov functions that describe the quark and antiquark inside the laser field.\\
To compute the hadronic decay width of $H^+$ in the presence of an EM field, we follow the same procedure as for the leptonic decay width. We get
\begin{equation}
\Gamma(H^{+}\rightarrow q \overline{q}')=\sum_{n=-\infty}^{+\infty}\Gamma^{n}(H^{+}\rightarrow q \overline{q}'),
\end{equation}
where $\Gamma^{n}(H^{+}\rightarrow q \overline{q}')$ is defined by
\begin{equation}
\Gamma^{n}(H^{+}\rightarrow q \overline{q}')=\frac{G_{F}\sqrt{2}|V_{qq'}|^{2}}{16(2\pi)^{2}Q}\int\frac{|\textbf{q}_{1}||\overline{\mathcal{M}^{n,h}_{fi}}|^{2}d\Omega_{q}}{Q+n\omega+\frac{\omega Q_{1}\cos(\theta)}{\sqrt{Q_{1}^{2}-m^{*2}_{q}}}\Big(n-\frac{e^{2}a^{2}}{2(k.k_{H^{+}})}\Big)}.
\end{equation}
The term $|\overline{\mathcal{M}^{n,h}_{fi}}|^{2}$ (the superscript $h$ stands for hadronic and $n$ is the number of photons) is reduced to trace calculation as follows:
\begin{equation}\label{trace2}
\begin{split}
|\overline{\mathcal{M}^{n,h}_{fi}}|^{2}=&\text{Tr}\Big[(\slashed{p}_{1}+m_{q})\Big\lbrace\Big(A+B\gamma^{5}\Big)b_{n}(z_{h})+\Big(C(p_{1})\slashed{a}_{1}\slashed{k}(A+B\gamma^{5})-C(p_{2})(A+B\gamma^{5})\slashed{k}\slashed{a}_{1}\Big)\\
&\times b_{1n}(z_{h})+\Big(C(p_{1})\slashed{a}_{2}\slashed{k}(A+B\gamma^{5})-C(p_{2})(A+B\gamma^{5})\slashed{k}\slashed{a}_{2}\Big)b_{2n}(z_{h})\Big\rbrace(\slashed{p}_{2}-m_{q'})\\
&\times\Big\lbrace\Big(A-B\gamma^{5}\Big) b^{*}_{n}(z_{h})+\Big(C(p_{1})(A-B\gamma^{5})\slashed{k}\slashed{a}_{1}-C(p_{2})\slashed{a}_{1}\slashed{k}(A-B\gamma^{5})\Big) b^{*}_{1n}(z_{h})\\
&+\Big(C(p_{1})(A-B\gamma^{5})\slashed{k}\slashed{a}_{2}-C(p_{2})\slashed{a}_{2}\slashed{k}(A-B\gamma^{5})\Big)b^{*}_{2n}(z_{h})\Big\rbrace\Big],
\end{split}
\end{equation}
where $z_{h}$ is the argument of Bessel functions. Note here that $C(p_{1})=\eta e/[2(k.p_1)]$ and $C(p_{2})=\eta' e/[2(k.p_2)]$, where the two factors $\eta=-2/3$ and $\eta'=1/3$ are, respectively, due to the fractional charge of up and down quarks. Regarding the sign, it should be noted that we considered $e<0$ as the charge of electron. The result of trace (\ref{trace2}) is included in appendix \ref{appendix}.
\subsection{Bosonic decay width of the charged Higgs in the presence of a laser field}\label{subbosonic}
The bosonic decay mode for the $H^+$ is depicted as follows:
\begin{equation}\label{bosonic_mode}
H^{+}(k_{H^{+}})\longrightarrow W^{+}(p_{W})+\Phi(p_{\Phi}),~~~~(\Phi\equiv h,H,A),
\end{equation}
with $h$, $H$ and $A$ being the light CP even neutral scalar, the heavy CP even neutral scalar and CP odd neutral pseudoscalar, respectively. The overall couplings between one gauge boson and two Higgs bosons in 2HDM of type II are defined as follows \cite{gunion2018}:
\begin{equation}
\begin{split}
 &\text{vertex}~H^{+}\text{-}W^{+}\text{-}h=\frac{-ig}{2}\cos(\beta-\alpha)(p_{W}^{\mu}+p_{h}^{\mu}),\\
 &\text{vertex}~H^{+}\text{-}W^{+}\text{-}H=\frac{-ig}{2}\sin(\beta-\alpha)(p_{W}^{\mu}+p_{H}^{\mu}),\\
 &\text{vertex}~H^{+}\text{-}W^{+}\text{-}A=\frac{-ig}{2}(p_{W}^{\mu}+p_{A}^{\mu}).
 \end{split}
 	\end{equation}
The lowest-order $S$-matrix element for the laser-assisted bosonic decay of the $H^+$ is as follows:
\begin{equation}
S_{fi}(H^{+}\rightarrow W^{+}\Phi)=\frac{-ig}{2}g_{{}_{HW\Phi}}\int d^{4}x\overline{\psi}_{\Phi}(x,p_{\Phi}) (p_{W}^{\mu}+p_{\Phi}^{\mu})\psi_{W}(x,p_{W})\overline{\phi}_{H}(x,k_{H^{+}}),
\end{equation}
where $g_{{}_{HW\Phi}}=\cos(\beta-\alpha)$, $g_{{}_{HW\Phi}}=\sin(\beta-\alpha)$ and $g_{{}_{HW\Phi}}=1$ correspond respectively to the decays  $H^{+}\rightarrow W^{+}h$, $H^{+}\rightarrow W^{+}H$ and $H^{+}\rightarrow W^{+}A$.  In order to take into account the interaction of the outgoing electrically charged $W^{+}$-boson (1-spin particle) with the EM field, we will describe it by the following wave function \cite{kurilin1999,obukhov2}:
\begin{equation}\label{wwave function}
\psi_{W}(x,p_{W})=\bigg[\textsl{g}_{\mu\nu}-\dfrac{e}{(k.p_{W})}\big(k_{\mu}A_{\nu}-k_{\nu}A_{\mu}\big)-\frac{e^{2}}{2(k.p_{W})^{2}}A^{2}k_{\mu}k_{\nu}\bigg]\frac{\varepsilon^{\nu}(p_{W},\lambda)}{\sqrt{2p^{0}_{W}V}}\times e^{iS(q_{W},x)},
\end{equation}
where $\textsl{g}_{\mu\nu}=\text{diag}(1,-1,-1,-1)$ is the metric tensor of Minkowski space, $\varepsilon^{\nu}(p,\lambda)$ is the $W^{+}$-boson polarization 4-vector such that $\sum_{\lambda=1}^{3}\varepsilon_{\mu}(p_{W},\lambda)\varepsilon_{\nu}^{*}(p_{W},\lambda)=-\textsl{g}_{\mu\nu}+p^{W}_{\mu}p^{W}_{\nu}/M_{W}^{2}$, and
\begin{equation}
S(q_{W},x)=-q_{W}.x+\dfrac{e(a_{1}.p_{W})}{k.p_{W}}\sin(\phi)-\dfrac{e(a_{2}.p_{W})}{k.p_{W}}\cos(\phi),
\end{equation}
with the dressed 4-momentum $q_{W}=(Q_{W},\textbf{q}_{W})$ and the effective mass $M_{W}^{*}$ that the boson $W^{+}$ acquires inside the EM field. For the final neutral scalar bosons, they are described by the following wave function:
\begin{equation}
\psi_{\Phi}(x,p_{\Phi})=\frac{1}{\sqrt{2p_{\Phi}^0V}}\times e^{-ip_{\Phi}.x},
\end{equation}
which obeys the Klein-Gordon equation for bosons with spin zero \cite{greinerqed}.\\
By following the same procedure as before, we obtain for the bosonic decay width
\begin{equation}
\Gamma(H^{+}\rightarrow W^{+}\Phi)=\sum^{+\infty}_{n=-\infty}\frac{\sqrt{2}G_{F}M^{2}_{W}}{8(2\pi)^{2}Q}g^{2}_{{}_{HW\Phi}}\int\frac{|\textbf{q}_{W}||\overline{\mathcal{M}^{n,b}_{fi}}|^{2}d\Omega_{W}}{Q+n\omega+\frac{\omega Q_{W}\cos(\theta)}{\sqrt{Q_{W}^{2}-M_{W}^{*2}}}\Big(n-\frac{e^{2}a^{2}}{2(k.k_{H^{+}})}\Big)},
\end{equation}
where
\begin{equation}
\begin{split}
|\overline{\mathcal{M}^{n,b}_{fi}}|^{2}=&\Big(-\textsl{g}^{\mu\nu}+\frac{p^{\mu}_{W}p^{\nu}_{W}}{M_{W}^{2}}\Big)\Big(p_{W}^{\mu}+p_{\Phi}^{\mu}\Big)\Big\lbrace \Big(\textsl{g}_{\mu\nu}-\dfrac{a^{2}e^{2}}{2(k.p_{W})}k_{\mu}k_{\nu}\Big) b_{n}(z_{b})+\frac{e}{k.p_{W}}\Big(k_{\mu}a_{1\nu}-k_{\nu}a_{1\mu}\Big) \\
&\times b_{1n}(z_{b})+\frac{e}{k.p_{W}}\Big(k_{\mu}a_{2\nu}-k_{\nu}a_{2\mu}\Big) b_{2n}(z_{b})\Big\rbrace\Big(p_{W}^{\nu}+p_{\Phi}^{\nu}\Big)\Big\lbrace \Big(\textsl{g}_{\nu\mu}-\dfrac{a^{2}e^{2}}{2(k.p_{W})}k_{\nu}k_{\mu}\Big) b^{*}_{n}(z_{b})\\
&+\frac{e}{k.p_{W}}\Big(k_{\nu}a_{1\mu}-k_{\mu}a_{1\nu}\Big) b^{*}_{1n}(z_{b})+\frac{e}{k.p_{W}}\Big(k_{\nu}a_{2\mu}-k_{\mu}a_{2\nu}\Big) b^{*}_{2n}(z_{b})\Big\rbrace,
\end{split}
\end{equation}
where $z_{b}$ is the Bessel function argument that appears in the calculation of the bosonic decay width of $H^+$. The quantity $|\overline{\mathcal{M}^{n,b}_{fi}}|^{2}$ (the superscript $b$ stands for bosonic) does not reduce here to the trace calculation since the bosonic decay process (\ref{bosonic_mode}) contains only scalar bosons which have no spin to sum over.
\section{Results and discussion}\label{sec:res}
\begin{figure}[t]
\centering
\includegraphics[scale=0.55]{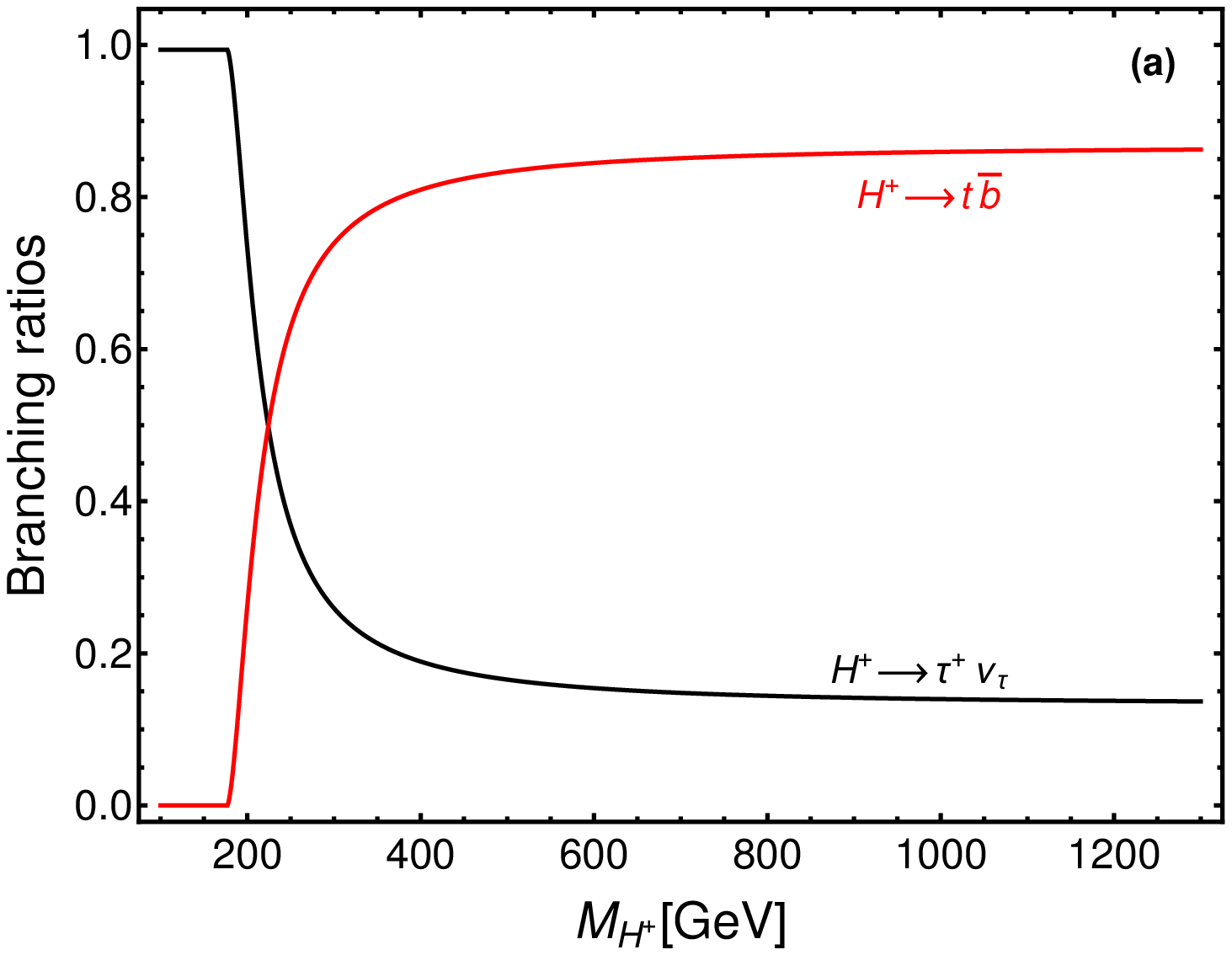}\hspace{0.2cm}
\includegraphics[scale=0.55]{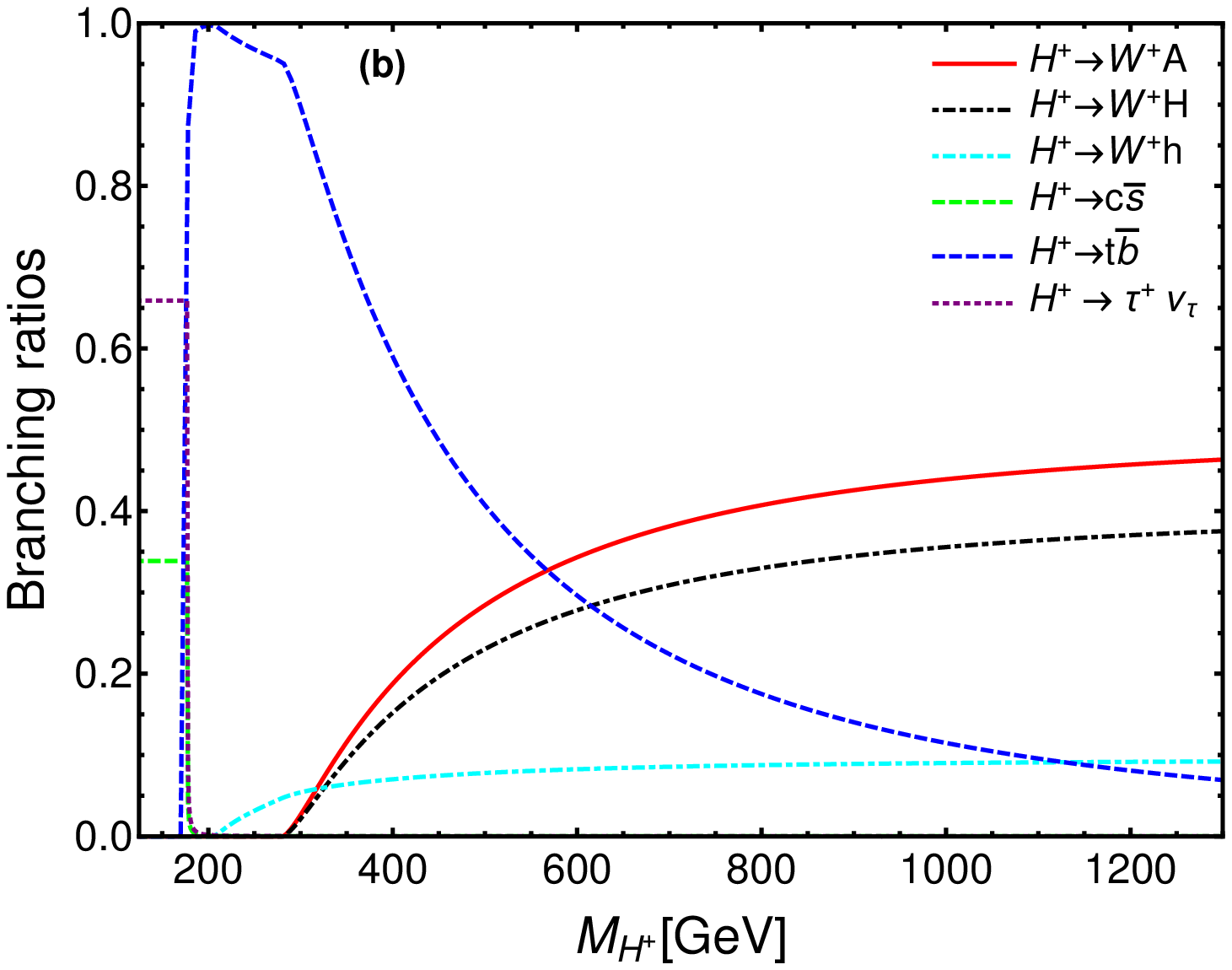}
\caption{Branching ratios of $H^+$ decay in 2HDM type II as a function of its mass, setting $\xi_0=0$ V/cm and $n=0$ (absence of laser field). The free parameters are \textbf{(a)} $M_{H^+}=M_H=M_A$, $m_h=125$ GeV, $\tan(\beta)=10$, $\sin(\beta-\alpha)=1$ (to be compared with \cite{ahmed2017}) and \textbf{(b)} $M_H=M_A=200$ GeV, $m_h=125$ GeV, $\tan(\beta)=1$, $\sin(\beta-\alpha)=0.9$ (to be compared with \cite{li2016}).}\label{fig3}
\end{figure}

This section will discuss the different numerical results obtained. It is quite appropriate to begin by examining the accuracy and consistency of our theoretical calculation. We have always been accustomed to comparing our theoretical expressions in the presence of a laser field with the corresponding ones in the absence of the laser field when applying the missed field limit, i.e., at $\xi_0=0$ V/cm and $n=0$ (no photons exchange). We have chosen to compare with the results previously presented in the following literature \cite{ahmed2017,li2016}. The thing which is known to be paid attention to when working beyond the standard model is the values of the free parameters. They must be carefully selected to respect current experimental limits. In our case, in addition to laser parameters ($\xi_0$ and $\omega$), the free model parameters are taken as follows: $M_{H^+}$, $M_{H}$, $M_{A}$, $m_{h}$, $\tan(\beta)$ and $\sin(\beta-\alpha)$. For $M_{H^+}$, we mention here that whenever we want to fix its value, we choose it in the range $570-800$ GeV, according to the recent study \cite{misiak2017}. To enforce that the energy-conservation law is always respected, we also note here that we have applied a condition (added in our computational program) on the effective mass of the charged Higgs, $M_{H^+}^*$, so that it always remains greater than the masses of the particles in the final state (on shell). Otherwise (off shell), the decay width is zero. The first reference \cite{ahmed2017}, to be compared with, adopted degenerate masses of the neutral Higgs bosons (i.e., $M_{H}=M_{A}=M_{H^+}$), and chose the remaining parameters as follows, $m_h=125$ GeV, $\tan(\beta)=10$ and $\sin(\beta-\alpha)=1$. In the latter conditions, all $H^+$ decay modes are absent except two, which are $H^+ \rightarrow t \bar{b}$ and $H^+ \rightarrow \tau^+ \nu_\tau$. If we also apply these conditions and make the missed field limit, we get the result shown in Fig.~\ref{fig3}\textbf{(a)}. The second comparison we made was with the result presented in \cite{li2016}. Our obtained result is shown in Fig.~\ref{fig3}\textbf{(b)}, where the free parameters are taken as follows: $M_{H}=M_{A}=200$ GeV, $m_h=125$ GeV, $\tan(\beta)=1$ and $\sin(\beta-\alpha)=0.9$. Figs.~\ref{fig3}\textbf{(a)} and \ref{fig3}\textbf{(b)} are in good agreement, respectively, with the result presented in \cite{ahmed2017} (see Fig.~8 therein) and \cite{li2016} (see Fig.~2 therein) if the same parameters are selected. These two figures are only intended to verify our theoretical calculation if it gives the result without laser in the limit of the vanishing field.
Because of the many decay modes of $H^+$, we decided to combine them into two, fermionic and bosonic. The first is the sum of the leptonic and hadronic widths, and the second is the bosonic width expressed in subsection~\ref{subbosonic}.
\begin{equation}
H^+\rightarrow \text{Fermions} = H^+\rightarrow \text{Hadr} + H^+\rightarrow \text{Lept},
\end{equation}
where
\begin{equation}
H^+\rightarrow \text{Hadr} = H^+\rightarrow u\bar{d},~c\bar{s},~t\bar{b},~~~\text{and} ~~~H^+\rightarrow \text{Lept} = H^+\rightarrow e^+ \nu_e,~\mu^+ \nu_\mu,~\tau^+ \nu_\tau,
\end{equation}
where other hadronic decay channels are greatly suppressed by CKM off-diagonal matrix elements. For the bosonic mode, we define
\begin{equation}
H^+\rightarrow \text{Bosons} = H^+\rightarrow W^+h,~ W^+H, ~W^+A.
\end{equation}
\begin{figure}[hbtp]
\centering
\includegraphics[scale=0.5]{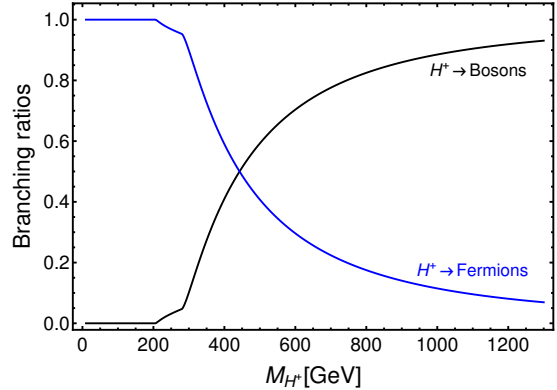}
\caption{Laser-free branching ratios of $H^+$ decay to bosonic and fermionic channels as a function of $M_{H^+}$. The free parameters are as in Fig.~\ref{fig3}\textbf{(b)}.}\label{fig4}
\end{figure}
The variations of the branching ratios for the fermionic and bosonic decay modes of $H^+$ are presented in Fig.~\ref{fig4} under the same conditions as in Fig.~\ref{fig3}\textbf{(b)}.\\
\begin{figure}[h]
 \centering
\includegraphics[scale=0.55]{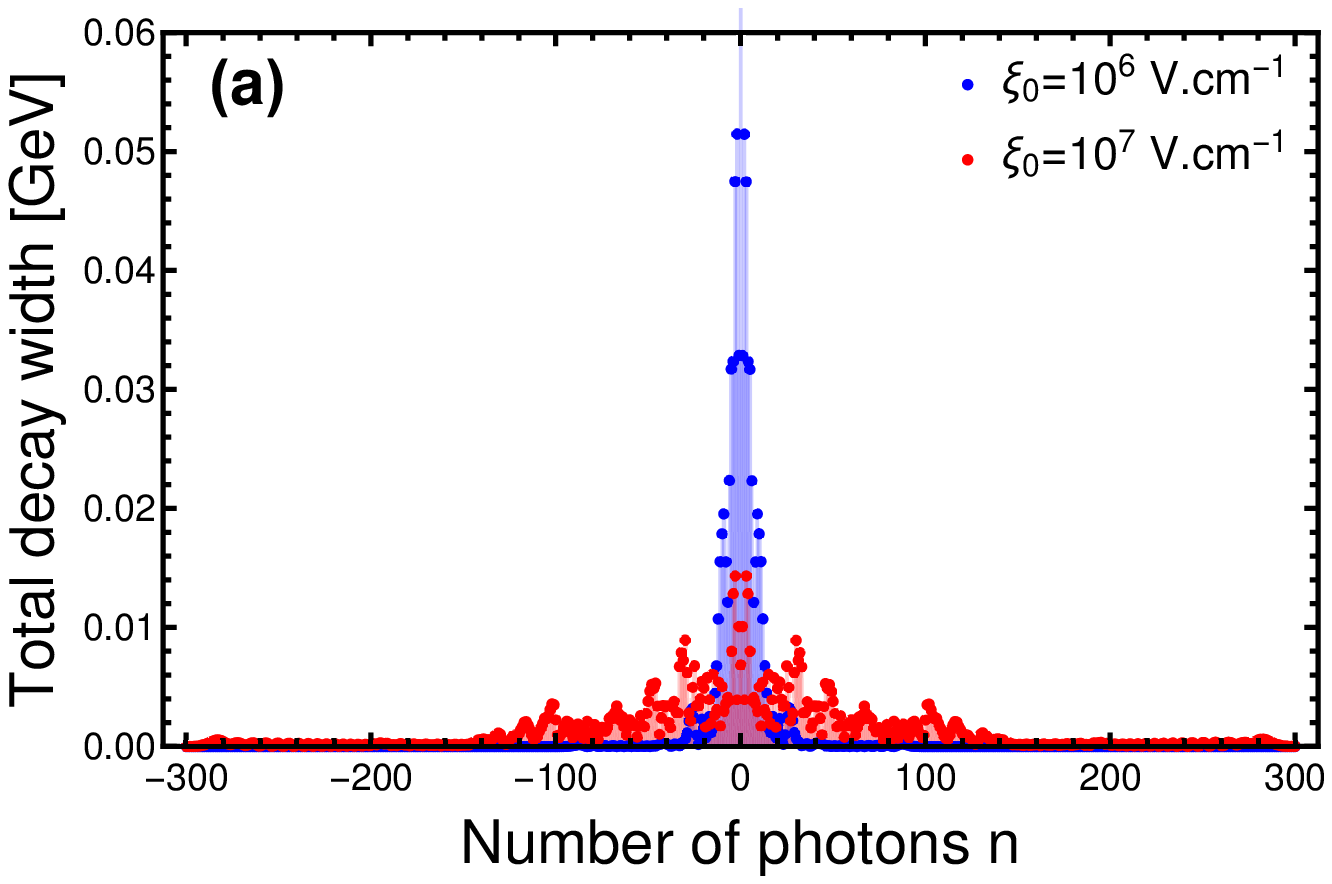}\hspace*{0.2cm}
\includegraphics[scale=0.55]{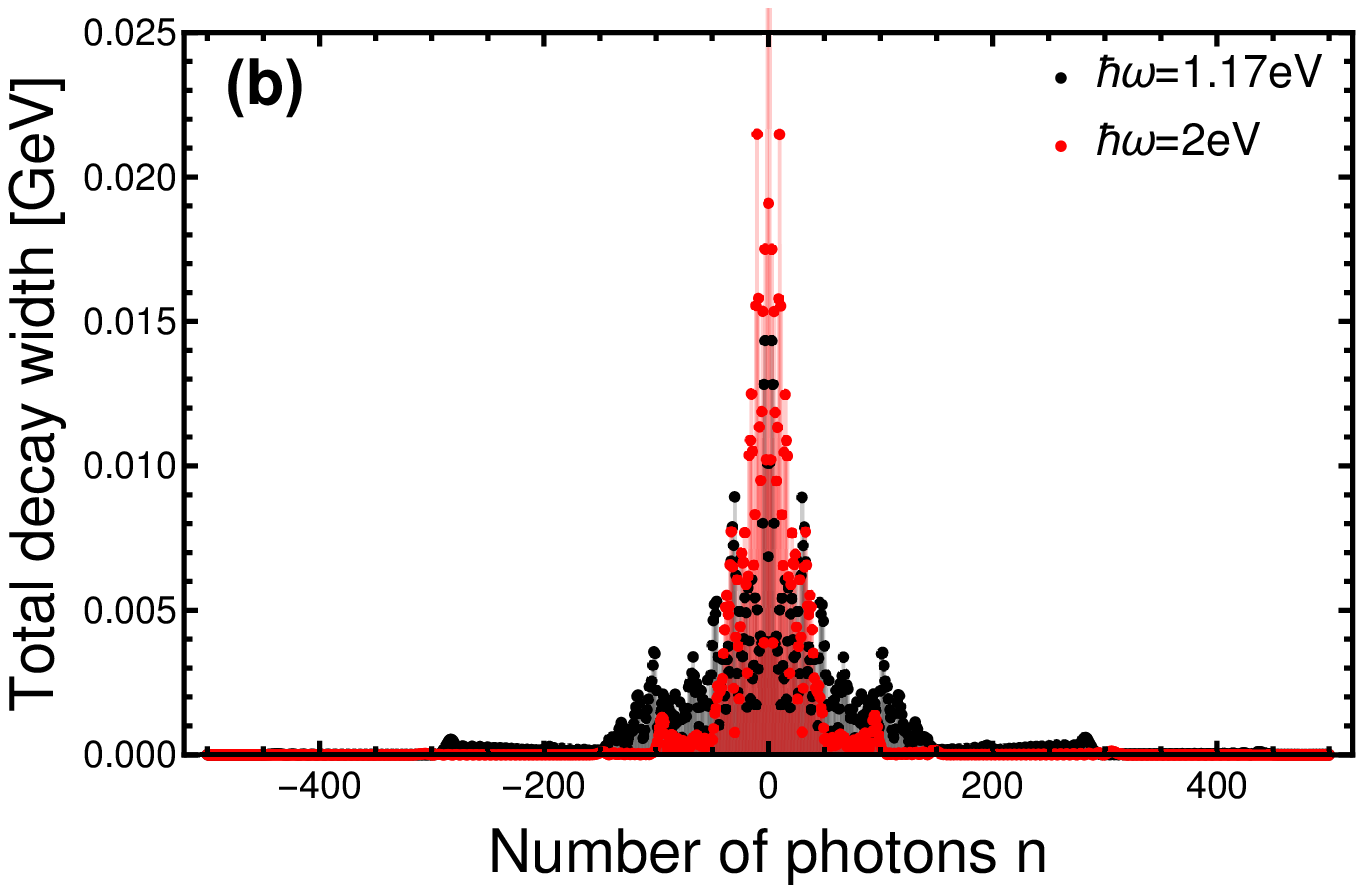}\par\vspace*{0.1cm}
\includegraphics[scale=0.55]{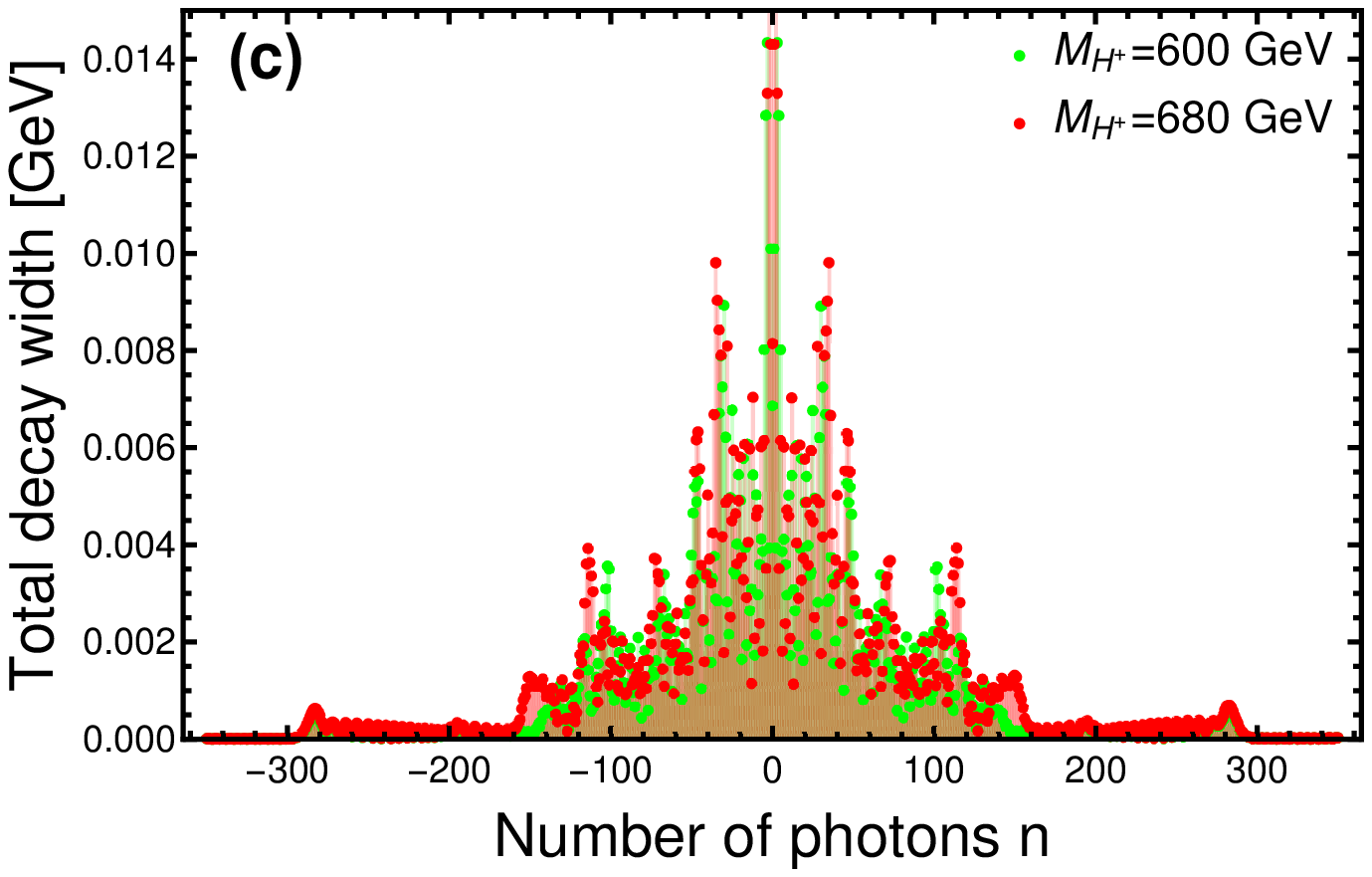}\hspace*{0.2cm}
\includegraphics[scale=0.55]{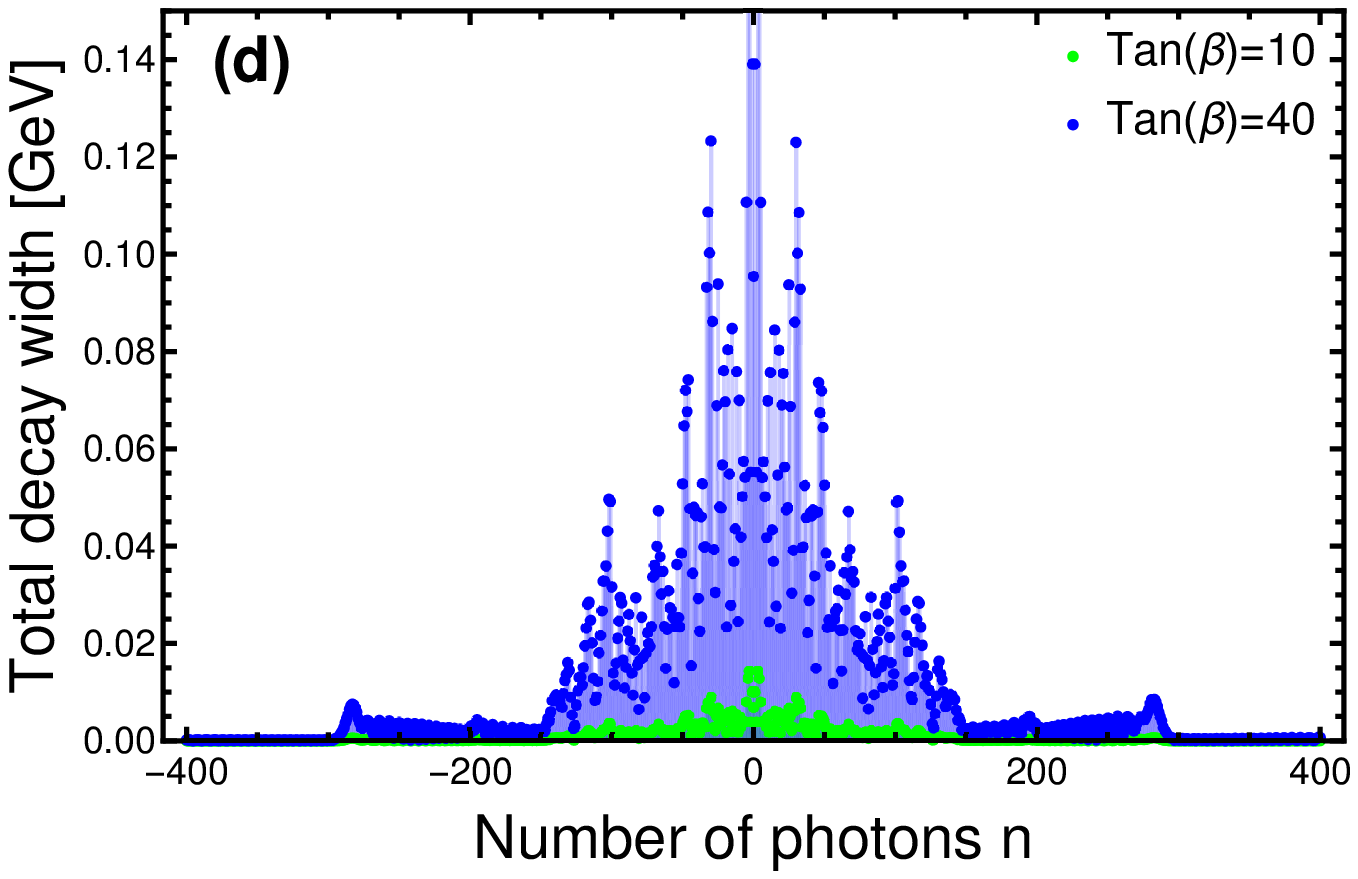}
\caption{Variation of the total decay width of $H^+$ in the presence of the laser versus the number of photons $n$. The free parameters are (unless otherwise stated) : $M_{H^+}=M_H=M_A=600$ GeV, $m_h=125$ GeV, $\sin(\beta-\alpha)=1$, $\tan(\beta)=10$, $\xi_0=10^7$ V/cm and $\hbar\omega=1.17$ eV.}\label{fig5}
\end{figure}
Given that the light charged Higgs boson has been excluded by the $B$-meson physics \cite{misiak2017,misiak2015}, we consider here the heavy charged Higgs regime. In this regime, and based on Fig.~\ref{fig3}\textbf{(a)}, we notice that channel $H^+\rightarrow t\bar{b}$ is the only dominant one respecting the conditions adopted there. In Figs.~\ref{fig3}\textbf{(b)} and \ref{fig4}, where the same free parameters have been chosen, we see that the two channels containing bosons in final state, exactly $H^+\rightarrow W^+H,~W^+A$, dominate in the heavy mass regime. \\
After checking the accuracy of our theoretical calculation by comparison with the recent literature, let us now consider the turn-on of the laser and discuss the decay of $H^+$ in the presence of a circularly polarized EM field. There are three parameters that characterize the laser when it is on, which are the field strength $\xi_0$, its frequency $\hbar\omega$, as well as the number of photons exchanged $n$. The interaction of the laser field with the decay system is therefore dependent on the modification of these parameters. The effect of the laser on $H^+$ decay is more reflected in the stimulated processes of absorption and emission of photons. To illustrate this phenomenon of multiphoton processes, we plot the variations of the total decay width $(\Gamma_\text{tot}=\Gamma(H^{+}\rightarrow \text{Fermions})+\Gamma(H^{+}\rightarrow \text{Bosons}))$ as a function of the number of photons $n$, as shown in Fig.~\ref{fig5}. The model parameters are chosen (unless otherwise stated) as follows: $M_{H^+}=M_H=M_A=600$ GeV, $m_h=125$ GeV, $\sin(\beta-\alpha)=1$, $\tan(\beta)=10$. For the laser parameters, we set $\xi_0=10^7$ V/cm and $\hbar\omega=1.17$ eV.
\\
Figure \ref{fig5}\textbf{(a)} displays the photon exchange phenomenon for two different field strengths, $10^6$ and $10^7$ V/cm. Each point corresponds to the value of the total decay width with respect to an integer number of photons $n$. The positive side represents emission and the negative one represents absorption. It is clear that the number of photons exchanged at $\xi_0=10^7$ V/cm is greater than that exchanged at $\xi_0=10^6$ V/cm. The increasing number of photons with higher field strength is evidence of the size of the laser effect on the decay system. Figure \ref{fig5}\textbf{(b)} shows the photon exchange process for two different frequencies. In contrast to Fig.~\ref{fig5}\textbf{(a)}, we see that the largest number of photons is exchanged at the lowest frequency. A small frequency means a pulse duration large enough to match the lifetime that the free charged Higgs boson is expected to live. Therefore, the effect is stronger at low frequencies than at high frequencies. In Fig.~\ref{fig5}\textbf{(c)}, it is obvious that the mass of the charged Higgs boson does not significantly affect the photon exchange, at least at $\xi_0=10^7$ V/cm. The charged Higgs boson with this heavy mass only feels the presence of the laser field at super intensities, unlike the leptons and hadrons in the final state. This is clearly seen in the effective mass (Eq.~(\ref{meffH})) values of $H^+$. For example, if we take the mass $M_{H^+}=600$ GeV, it remains the same in the presence of the laser until $\xi_0=10^{13}$ V/cm. At super intensities (e.g., $\xi_0=10^{16}$ V/cm and $\omega=1.17$ eV), the effective mass becomes $M^*_{H^+}=623.253$ GeV. This means that the laser made $H^+$ gain an additional $23.253$ GeV of mass. Figure \ref{fig5}\textbf{(d)} indicates that large values of $\tan(\beta)$ induce the exchange of photons between the laser and the decay system.
\begin{table}
\caption{Numerical values of branching ratios for $H^+$ decay as a function of the laser field strength, with $M_{H^+}=620$ GeV and $\hbar\omega=0.117$ eV. The model parameters are as in figure \ref{fig3}\textbf{(b)}.}\label{tab1}
\begin{center}
\begin{tabular}{ccc}
\toprule
$\xi_{0}$ $[\text{V/cm}]$ & BR$(H^{+}\rightarrow \text{Fermions})$  & BR$(H^{+}\rightarrow \text{Bosons})$  \\ \hline
 $10$       & $0.2793$        &  $0.7206$   \\
 $10^{2}$   & $0.2793$       &   $0.7206$   \\
 $10^{3}$   & $0.27935$       &   $0.7206$   \\
 $10^{4}$   & $0.3587$       &    $0.64124$   \\
 $10^{5}$   & $0.5172$       &   $0.4827$   \\
 $10^{6}$   & $0.5845$       &   $0.4154$  \\
 $10^{7}$   & $0.57840$       & $0.4216$ \\
 $10^{8}$   & $0.5742$       &  $0.4257$  \\
 $10^{9}$   & $0.58890$      &  $0.4110$  \\
 $10^{10}$   & $0.5667$     &  $0.4332$  \\
 $10^{11}$   & $0.1300$     &  $0.8699$  \\
 $10^{12}$   & $0.000026$     &  $0.9999$  \\
\toprule
\end{tabular}
\end{center}
\end{table}
After discussing the effect of the laser on the total decay width and the photon exchange process, let us now turn to the analysis of the laser effect on the branching ratios. We shall do so in both choices of free parameters, as in Figs.~\ref{fig3}\textbf{(a)} and \ref{fig3}\textbf{(b)}. We start with the conditions chosen in Fig.~\ref{fig3}\textbf{(b)}. Table \ref{tab1} gives the numerical values of the bosonic and fermionic branching ratios in terms of field strength. The free parameters are chosen as in Fig.~\ref{fig3}\textbf{(b)}, with $M_{H^+}=620$ GeV. Through this table, it is clear that the low strengths of the laser field $[10-10^4~\text{V/cm}]$ did not affect the branching ratios, since the bosonic mode continued to be dominant, as in the absence of the laser. However, the laser effect starts to appear at high and medium field strengths $[10^5-10^{10}~\text{V/cm}]$. The laser has now enhanced the fermionic mode over the bosonic one. At field strengths of $10^{11}$ and $10^{12}$ V/cm (ultrastrong field), the laser made a dramatic and sudden change. It blocks the fermionic mode to fully open the bosonic one, which becomes the only one allowed at $10^{12}$ V/cm of almost $100\%$. Note that the two branching ratios listed in the table compensate for each other because of the probabilistic meaning they have. This means that their sum is 1 (or $100\%$).
\begin{figure}[hbtp]
\centering
\includegraphics[scale=0.55]{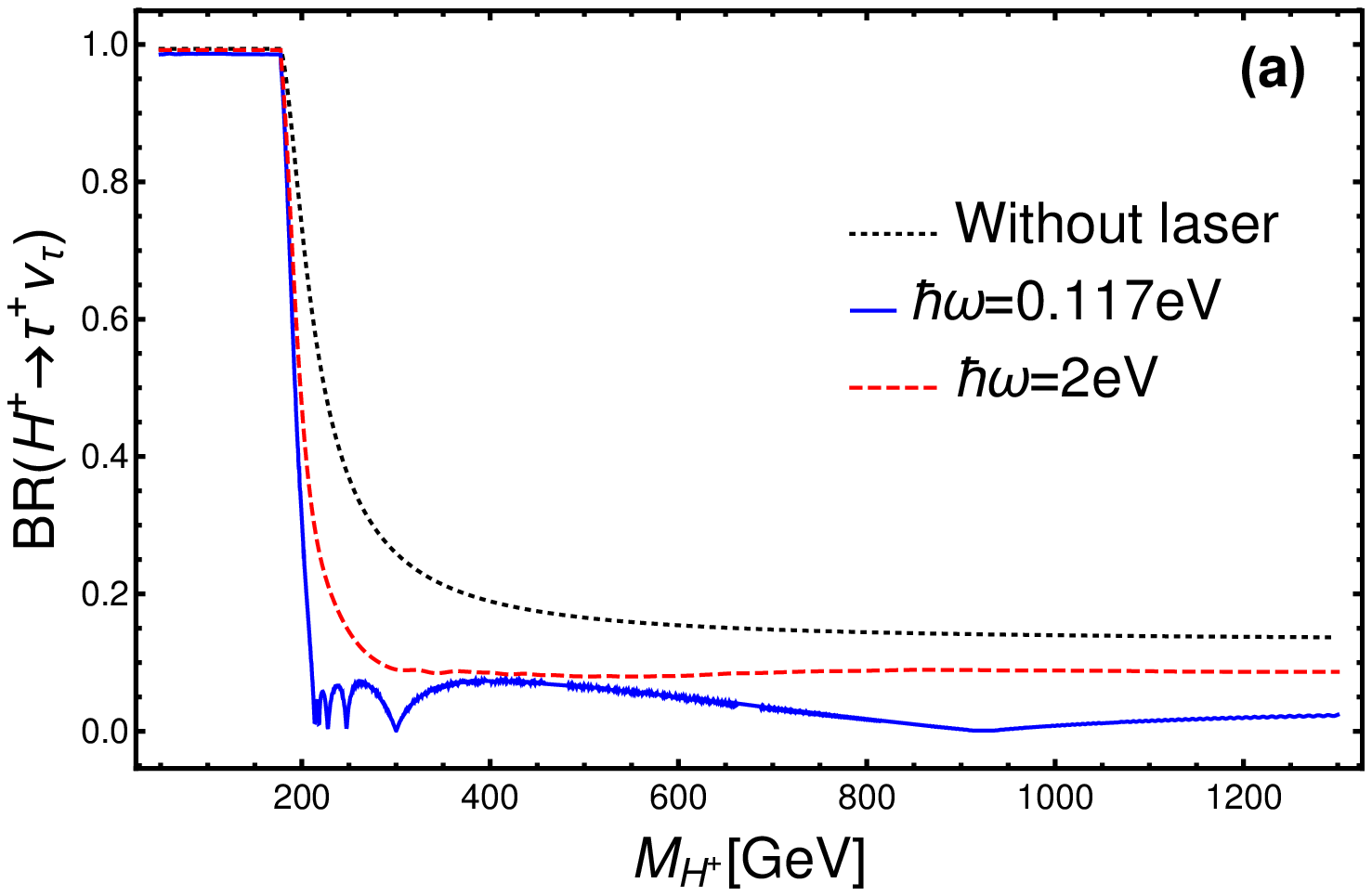}\hspace*{0.2cm}
\includegraphics[scale=0.55]{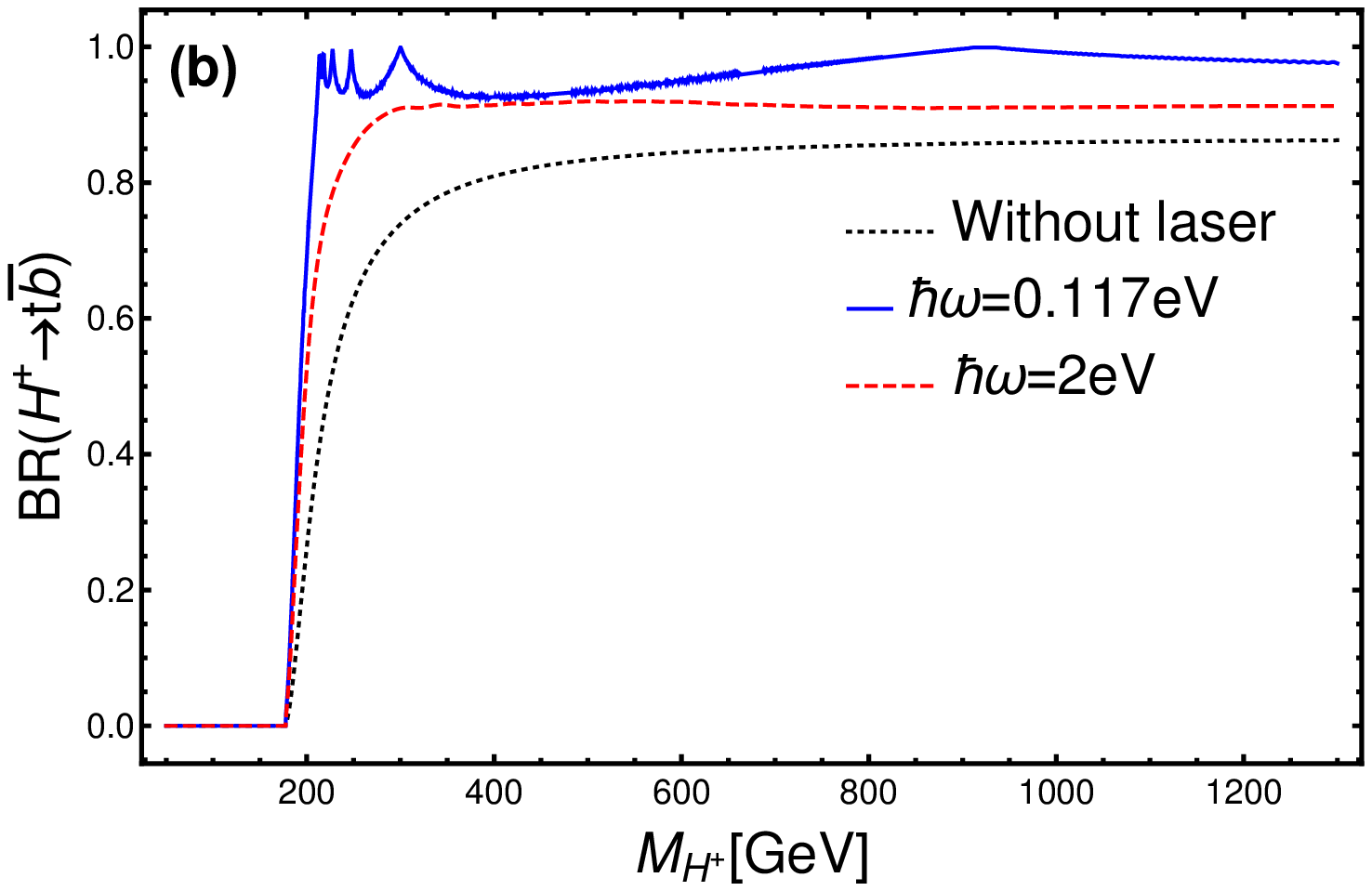}
\caption{Variations of branching ratios for laser-assisted $H^+$ decay as a function of $M_{H^+}$ for two different frequencies at $\xi_0=10^7$ V/cm. The model parameters are chosen as in Fig.~\ref{fig3}\textbf{(a)}.}\label{fig6}
\end{figure}\\
Under the same conditions as in Fig.~\ref{fig3}\textbf{(a)}, Fig.~\ref{fig6} displays the variations of branching ratios as a function of $M_{H^+}$ for two different frequencies $(0.117~\text{and}~2~\text{eV})$ at the field strength $\xi_0=10^7$ V/cm. In this case, as mentioned before, among all available channels, only $t\bar{b}$ and $\tau^+\nu_\tau$ channels remain open. As can be seen in this figure, the laser field strength of $10^7$ V/cm had no significant effect on the branching ratios. Indeed, the $t\bar{b}$ channel was initially dominant over $\tau^+\nu_\tau$ one in the absence of laser at the heavy Higgs masses. In the presence of a laser field of strength $10^7$ V/cm, there was only further enhancement of $t\bar{b}$ channel and suppression of the other depending on the frequency used. The effect of a laser with a lower frequency is significant compared to that with a higher frequency.
\begin{figure}[hbtp]
\centering
\includegraphics[scale=0.6]{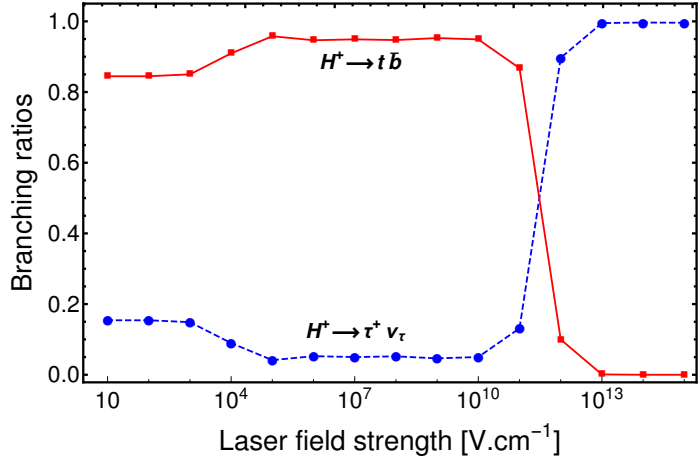}
\caption{Variations of branching ratios for laser-assisted $H^+$ decay as a function of the field strength. The model parameters are chosen as in Fig.~\ref{fig3}\textbf{(a)}, with $M_{H^+}=600$ GeV. The laser frequency is $\hbar\omega=0.117$ eV.}\label{fig7}
\end{figure}
Now, let us increase the laser field strength and see what happens in the ultra-high intensities regime. Figure \ref{fig7} shows the branching ratio changes versus the field strength under the same conditions as in Fig.~\ref{fig3}\textbf{(a)}. As can be seen in Fig~\ref{fig7}, there is no significant effect of the laser field on the branching ratios in the field strength range of $10$ to $10^{11}$ V/cm. But, once this range is exceeded, the laser makes a significant contribution to strongly enhancing $\tau^+\nu_\tau$ channel and accordingly reducing $t\bar{b}$. Although these super intensities are not yet available experimentally in laboratories, there is currently tremendous work to develop infrastructure for powerful laser sources, and it is only a matter of time before higher intensities are reached in the near future. We have carefully stopped at $10^{15}$ V/cm to avoid pair creation that occurs around $10^{16}$ V/cm (Schwinger limit) \cite{schwinger1,schwinger2,schwinger3}. This result, which is the change in branching ratios due to the laser, is very interesting and requires experimental investigation.
\begin{figure}[hbtp]
\centering
\includegraphics[scale=0.6]{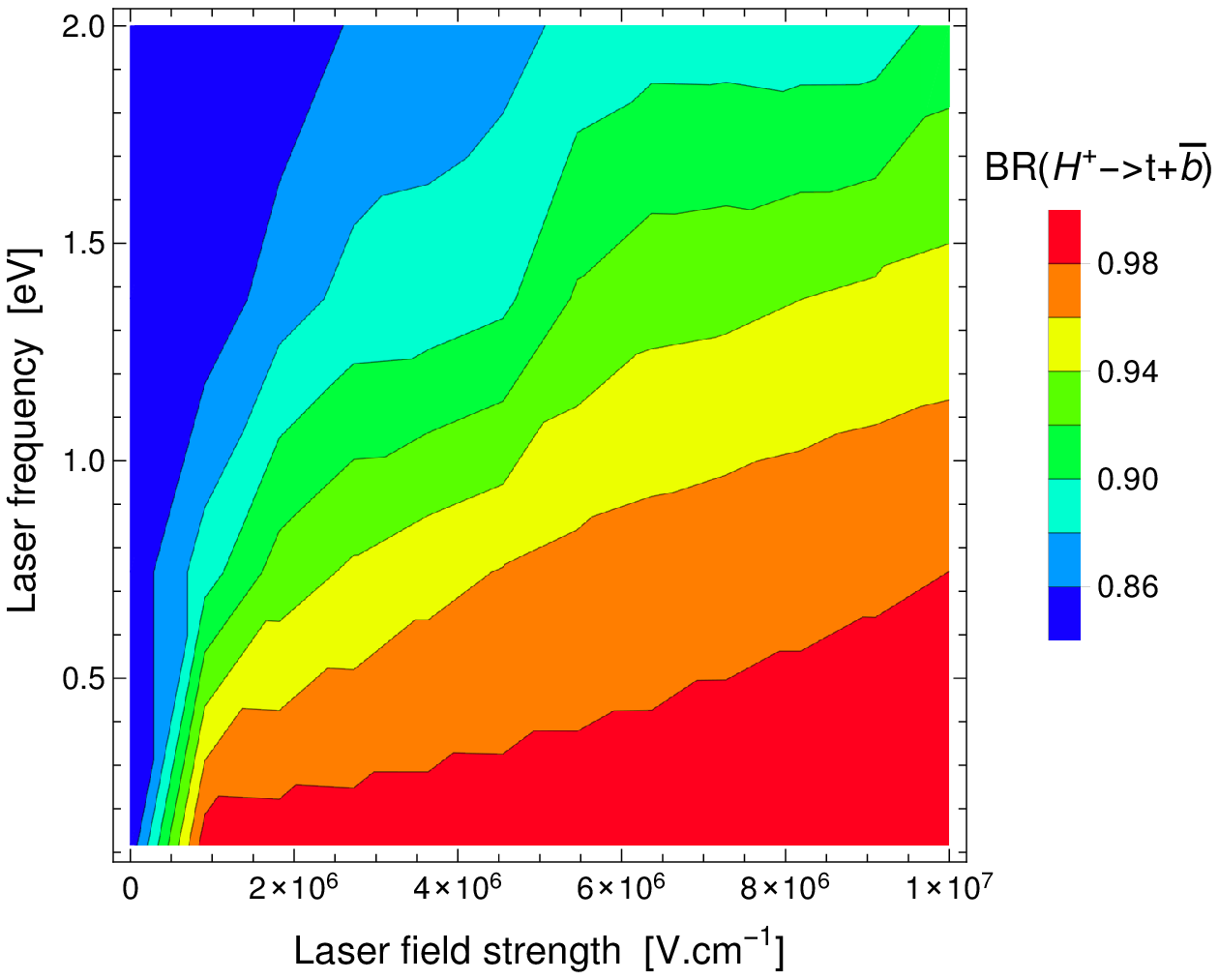}\hspace*{0.2cm}
\includegraphics[scale=0.6]{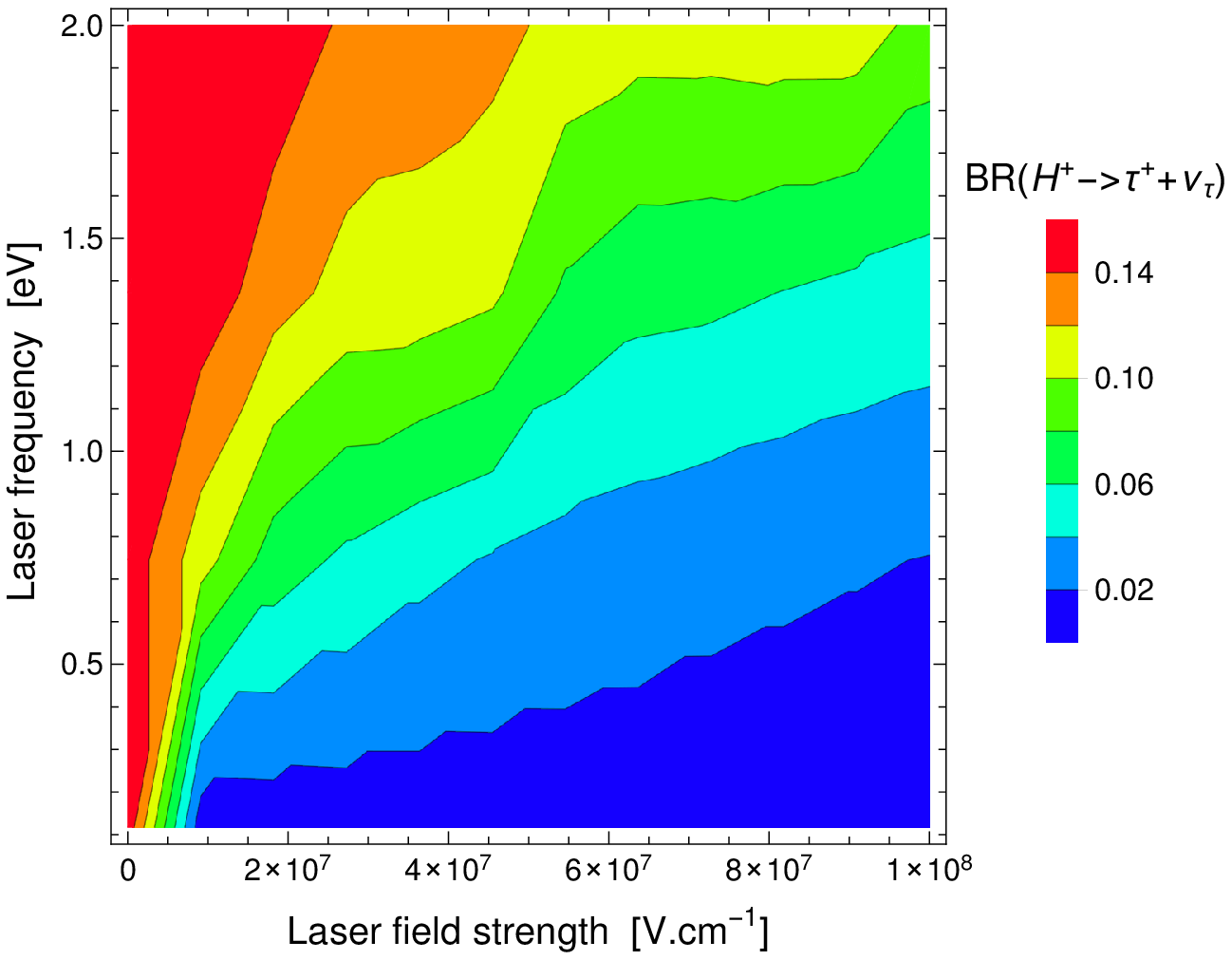}
\caption{Contour plot of branching ratios for laser-assisted $H^+$ decay as a function of the field strength and frequency. The model parameters are chosen as in Fig.~\ref{fig3}\textbf{(a)}, with $M_{H^+}=620$ GeV.}\label{fig8}
\end{figure}\\
To give a clearer picture, we present in Fig.~\ref{fig8} a contour plot of the branching ratios, where their changes are highlighted in terms of field strength and frequency together. One can observe how the branching ratios change as the laser field strength increases at a specific frequency or vice versa. The effect of the laser becomes very significant at high field strengths and low frequencies. This confirms the previous discussion in Figs.~\ref{fig5}\textbf{(a)} and \ref{fig5}\textbf{(b)} about the effect of laser field strength and frequency on the photon exchange process. Moreover, we can see how the two contours are complementary and superposable to each other.
\section{Conclusion}\label{sec:conclusion}
In summary, we provided theoretical evidence that the branching ratios of charged Higgs decay can be modified in a large range by applying an appropriate laser field. Using $S$-matrix method, we have performed analytical calculations for charged Higgs decay in the presence of a circularly polarized radiation field. Our main result is shown in Fig.~\ref{fig7}, where the chosen free parameters allow only two channels, $H^+\rightarrow \tau^+\nu_\tau$ and $H^+\rightarrow t\bar{b}$. It is revealed that $\tau^+\nu_\tau$ channel, which is the lowest in the absence of a laser, becomes totally dominant in the region of superstrong fields $[10^{12}-10^{13}~\text{V/cm}]$. The fact that an appropriate laser field can modify branching ratios is extremely important, especially when dealing with a yet undiscovered particle. Recently, we found that the laser has an unprecedented effect on the branching ratios of vector bosons $W^-$ and $Z$ \cite{wdecay,zdecay}. Without forgetting that these results remain purely theoretical and therefore require experimental verification.  It is time to redouble our efforts to take advantage of laser technology and implement it into large colliders in parallel with its continuous progress since the 1960s. The intense electromagnetic environment, combined with conventional acceleration, could be a prospective way to increase the collision energy required for detection of the charged Higgs boson.
\appendix
\section{Calculation of traces}\label{appendix}
We present here the results of traces calculation appeared in Eqs.~(\ref{trace1}) and (\ref{trace2}). For the leptonic decay mode calculation, Eq.~(\ref{trace1}) reads
\begin{equation}
\begin{split}
|\overline{\mathcal{M}_{fi}^n}|^2 =&\frac{1}{(k.p_{1})^{2}}\Big[2 e \Big(J_{n-1}(z_{\ell}) - J_{n+1}(z_{\ell})\Big)\Big(e (k.p_{2})\Big(J_{n-1}(z_{\ell}) + J_{n+1}(z_{\ell})\Big)\epsilon(a_{1},a_{2},k,p_{1})\\
&+2 J_{n}(z_{\ell}) (k.p_{1}) \Big(\cos(\phi_{0})\epsilon(a_{2},k,p_{1},p_{2})-\sin(\phi_{0})\epsilon(a_{1},k,p_{1},p_{2})\Big)\Big)+2 (k.p_{1})\\
&\times \Big(-a^2 e^2 (k.p_{2})\Big(J^{2}_{n-1}(z_{\ell})+J^{2}_{n+1}(z_{\ell})\Big)+2 e J_{n}(z_{\ell}) \Big(J_{n-1}(z_{\ell}) + J_{n+1}(z_{\ell})\Big)\cos(\phi_{0})\\
&\times\Big((a_{1}.p_{2})(k.p_{1})-(a_{1}.p_{1})(k.p_{2})\Big)+4 J^{2}_{n}(z_{\ell})(k.p_{1})(p_{1}.p_{2})\Big)\Big],
\end{split}
\end{equation}
where for all 4-vectors $a, b, c$ and $d$, $\epsilon(a,b,c,d)=\epsilon_{\mu\nu\rho\sigma}a^{\mu}b^{\nu}c^{\rho}d^{\sigma}$. For the Levi-Civita tensor, $\epsilon_{\mu\nu\rho\sigma}$, the convention $\epsilon_{0123}=+1$ is used.\\
For the hadronic decay mode calculation, Eq.~(\ref{trace2}) yields
\begin{equation}
\begin{split}
|\overline{\mathcal{M}^{n,h}_{fi}}|^2 =&\Delta_{1}|b_{n}(z_{h})|^{2}+\Delta_{2}|b_{1n}(z_{h})|^{2}+\Delta_{3}|b_{2n}(z_{h})|^{2}+\Delta_{4}b_{1n}(z_{h})b^{*}_{2n}(z_{h})+\Delta_{5}b_{2n}(z_{h})b^{*}_{1n}(z_{h})\\
&+\Delta_{6}b_{n}(z_{h})b^{*}_{1n}(z_{h})+\Delta_{7}b_{1n}(z_{h})b^{*}_{n}(z_{h})+\Delta_{8}b_{2n}(z_{h})b^{*}_{n}(z_{h})+\Delta_{9}b_{n}(z_{h})b^{*}_{2n}(z_{h}),\\
\end{split}
\end{equation}
where the coefficients $\Delta_{1}$ to $\Delta_{9}$ are expressed by
\begin{equation}
\Delta_{1}=4 \Big(A^2 (-m_{q} m_{q'} + (p_{1}.p_{2})) + B^2 (m_{q} m_{q'} + (p_{1}.p_{2}))\Big),
\end{equation}
\begin{equation}
\Delta_{2}=-\frac{2 a^2 (A^2 + B^2) e^2 ((k.p_{2})\eta + (k.p_{1}) \eta')^2}{(k.p_{1})(k.p_{2})},
\end{equation}
\begin{equation}
\Delta_{3}=-\frac{2 a^2 (A^2 + B^2) e^2 ((k.p_{2})\eta + (k.p_{1})\eta')^2}{(k.p_{1})(k.p_{2})},
\end{equation}
\begin{equation}
\Delta_{4}=\frac{2 i A~B e^2 \Big((k.p_{2})\eta + (k.p_{1}) \eta'\Big)^2\Big((k.p_{2})\epsilon(a_{1},a_{2},k,p_{1})+(k.p_{1})\epsilon(a_{1},a_{2},k,p_{2})\Big) }{(k.p_{1})^2 (k.p_{2})^2},
\end{equation}
\begin{equation}
\Delta_{5}=-\frac{2 i A~B e^2 \Big((k.p_{2})\eta + (k.p_{1}) \eta'\Big)^2\Big((k.p_{2})\epsilon(a_{1},a_{2},k,p_{1})+(k.p_{1})\epsilon(a_{1},a_{2},k,p_{2})\Big)}{(k.p_{1})^2(k.p_{2})^2},
\end{equation}
\begin{equation}
\Delta_{6}=\frac{2 e \Big((k.p_{2}) \eta + (k.p_{1})\eta'\Big)\Big(\text{-}\Big(A^2 + B^2\Big)\Big((a_{1}.p_{2})(k.p_{1}) - (a_{1}.p_{1})(k.p_{2})\Big)-2 i A~B\epsilon(a_{1},k,p_{1},p_{2})\Big)}{(k.p_{1})(k.p_{2})},
\end{equation}
\begin{equation}
\Delta_{7}=\frac{2 e \Big((k.p_{2}) \eta + (k.p_{1})\eta'\Big)\Big(\text{-}\Big(A^2 + B^2\Big)\Big((a_{1}.p_{2})(k.p_{1}) - (a_{1}.p_{1})(k.p_{2})\Big)+2 i A~B\epsilon(a_{1},k,p_{1},p_{2})\Big)}{(k.p_{1})(k.p_{2})},
\end{equation}
\begin{equation}
\Delta_{8}=\frac{4 i A~B e \Big((k.p_{2}) \eta + (k.p_{1}) \eta'\Big)\epsilon(a_{2},k,p_{1},p_{2})}{(k.p_{1})(k.p_{2})},
\end{equation}
\begin{equation}
\Delta_{9}=-\frac{4 i A~B e \Big((k.p_{2}) \eta + (k.p_{1}) \eta'\Big)\epsilon(a_{2},k,p_{1},p_{2})}{(k.p_{1})(k.p_{2})}.
\end{equation}

\end{document}